\documentclass[twocolumn,showpacs,amsmath,amssymb,nofootinbib,pra,superscriptaddress]{revtex4-1}

\usepackage{comment}
\usepackage{graphicx} 
\usepackage{xcolor}

\usepackage[%frenchlinks=true,
colorlinks=true,
    linkcolor={red!50!black},
    citecolor={blue!50!black},
    urlcolor={blue!80!black}]{hyperref}

\DeclareMathAlphabet\mathbfcal{OMS}{cmsy}{b}{n}

\begin{document}

\title{Analytical approach to the Bose polaron problem in one dimension}

\author{A.~G. \surname{Volosniev}}
\affiliation{Institut f{\"u}r Kernphysik, Technische Universit{\"a}t Darmstadt, 64289 Darmstadt, Germany}
\author{H.-W. \surname{Hammer}}
\affiliation{Institut f{\"u}r Kernphysik, Technische Universit{\"a}t Darmstadt, 64289 Darmstadt, Germany}
\affiliation{
ExtreMe Matter Institute EMMI, GSI Helmholtzzentrum f{\"u}r Schwerionenforschung GmbH, 64291 Darmstadt, Germany}

\begin{abstract}
  We discuss the ground state properties of a one-dimensional bosonic system doped with an impurity (the so-called Bose polaron problem).  We introduce a formalism that allows us to calculate analytically the thermodynamic zero-temperature properties of this system with weak and moderate boson-boson interaction strengths for any
  boson-impurity interaction. Our approach is
  validated by comparing to exact quantum Monte Carlo calculations.
  In addition, we test the method in finite size systems using numerical results
  based upon the similarity renormalization group.   
   We argue that the introduced approach provides a simple analytical tool for studies of strongly interacting impurity problems in  one dimension. 
\end{abstract}

%\pacs{67.85.-d, %Ultracold gases, trapped gases
%67.85.Pq, %Mixtures of Bose and Fermi gases
%?
%}

\maketitle

The weakly-interacting Bose gas is a beautiful model system~\cite{landau9},
which is often used to study emergent many-body phenomena such as superfluidity,
Bose-Einstein condensation, and topologically non-trivial many-body excitations like solitons and vortices. The basic properties
of this model are well understood theoretically and tested experimentally
(see, e.g., Refs.~\cite{dalfovo,Andersen:2003qj}). However, some important questions still remain open.
One of them concerns the reaction of a Bose gas to a mobile impurity
particle, which is usually referred to as the Bose polaron problem, in analogy to the  polaron
studied by Landau and Pekar~\cite{landaupekar}. 
Polaron problems are among the simplest problems exhibiting non-trivial many-body effects that shed light on the interplay of
one- and many-body physics.
However, the fate of the impurity in a gas is not of only formal
interest. Properties of many systems in condensed matter physics can be understood by studying a single particle interacting with a reservoir.
Prominent examples are given by a single $^3$He atom in liquid $^4$He~\cite{dobbs2000} and an electron in an ionic crystal (often described as a particle interacting with a Bose field of ion vibrations), see \cite{devresee2009} and references therein. 
The apparent simplicity of these problems is misleading, as to date they  resist a full theoretical solution. Fortunately,
experiments with cold atomic gases, realizing the idea of a quantum simulator \cite{feynman1982}, open up the possibility to create and study the Bose polaron~\cite{kohl2009, schmid2010, kohl2010, widera2012, catani2012, levinsen2016, cornell2016, meinert2016} in a laboratory.  
 This intriguing possibility motivated a flurry of recent theoretical works on this problem~\cite{pitaevski2004,blume2006,timmermans2006,bruderer2008,devreese2009,schecter2012,schmidt2013, devreese2015,grusdt2015, giorgini2015a, volosniev2015, grusdt2015a, levinsen2015, levinsen2015a, giorgini2016, giorgini2016a, schecter2016, naidon2016,cui2017,vacarchuk2017}.

One-dimensional (1D) systems are of special interest in this context, because strongly interacting bosons
in 1D fermionize~\cite{girardeau1960}. This phenomenon simplifies the analysis. For example, if all the masses in the system are
identical, the Bose polaron problem is exactly solvable~\cite{mcguire1965}. This is also true if the impurity is infinitely heavy~\cite{giorgini2016}. Therefore, a solution of the problem
for weak and moderate boson-boson interaction is enough to complete the picture for all interaction
strengths.  However, theoretical approaches face challenges in describing these parameter regimes if the boson-impurity interactions are strong~\cite{grusdt2017}. In this case accurate results can be obtained only numerically using Monte-Carlo methods~\cite{giorgini2016,grusdt2017}, and analytical calculations that can unravel underlying physics and correlations are highly desirable. 
In this Rapid Communication,
we introduce a possible theoretical formalism for performing such calculations. Our approach is well-suited for
studying the energy and structural properties of the Bose polaron problem, and for investigating systems with finite
number of particles.  To illustrate this statement, we present analytical expressions for the ground state energy and contact parameter in
the thermodynamic limit and show that they agree with the recent numerical results based upon the quantum Monte Carlo
method~\cite{giorgini2016}. Furthermore, we test our method for finite systems using results of a numerical similarity
renormalization group method as a benchmark.

\section{Formulation} 

We study a system that consists of an impurity of mass $m$, and $N$ bosons of mass $M$ on a ring of length $L$.
This system is described by the Schr{\"o}dinger equation $H\Psi=\varepsilon\Psi$ with the Hamiltonian
\begin{align}
H=
-\frac{\hbar^2}{2m}\frac{\partial^2}{\partial y^2}-\frac{\hbar^2}{2M}\sum_{i=1}^N\frac{\partial^2}{\partial x_i^2}+ V_{ib}+
V(\{x_i\}), 
\label{eq:ham1}
\end{align} 
where $y$ is the position of the impurity, $x_i$ is the position of the $i$th boson, and the boson-boson interaction is
given by
$V(\{x_i\})=g\sum\delta(x_i-x_j)$,  where $g\geq 0$ to have a well-defined thermodynamic limit~\cite{mcguire1964}. The bosons
interact with the impurity via the potential $V_{ib}$, which we write as $V_{ib}(x)=c\sum\delta(x_i-y)$, with $c\geq 0$. Note that the presented approach can be generalized straightforwardly to systems with finite range potentials. We do not pursue this possibility here, and only note that it will allow one to test the accuracy of the delta-function approximation for the interaction of an atom (ion) with a boson. 
For later convenience, we set $\hbar=M=1$ in what follows.  

We are interested in the ground state properties of the Hamiltonian $H$, and in particular, in the quantity $\epsilon=\varepsilon_{gr}(c)-\varepsilon_{gr}(c=0)$, to which we will refer as the energy of the impurity. Since the  Schr{\"o}dinger equation is analytically solvable for $c=0$~\cite{lieb1963}, the knowledge of $\epsilon$ gives us directly $\varepsilon_{gr}(c)$.
To find $\epsilon$, we use the equation 
\begin{align}
-\frac{1}{2}\sum_{i=1}^N \frac{\partial^2 \Phi}{\partial z_{i}^2}-\frac{1}{2 m}\left(\sum_{i=1}^N \frac{\partial}{\partial z_i}\right)^2\Phi  + V(\{z_i\})\Phi  =\varepsilon_{gr}\Phi,
\label{eq:imp_schr_1}
\end{align}
for the function $\Phi(z_1, .., z_N)$ assuming that $z_i\in (0,L)$. The boundary conditions are taken as
\begin{equation}
\frac{\partial \Phi}{\partial z_i}\bigg|^{z_i=0^+}_{z_i=L^-}=2c\kappa \Phi|_{z_i=0}, \; \Phi|_{z_i=0}=\Phi|_{z_i=L},
\label{eq:bound_cond_initial}
\end{equation}
where $\kappa=m/(1+m)$ is the reduced mass of the impurity and a boson. 
The notation $z_i=p^{\pm}$ means that the derivative is taken at the point
$z_i=p\pm o$ with $o > 0$ and the limit $o\to 0$ is taken
afterwards. 
The function $\Phi$ can be used to solve the original problem if for every ordering of particles, e.g., $0<y<x_1<...<x_{N}<L$, the following prescription is applied: $z_i=L\theta(y-x_i)+(x_i-y)$ where $\theta$ is the Heavyside step function, i.e., $\theta(x>1)=1$ and zero otherwise. Therefore, Equation~(\ref{eq:imp_schr_1}) is the Schr{\"o}dinger equation (with zero center-of-mass motion) in which all distances are measured with respect to the impurity. Its second term contains information about the kinetic energy of the impurity.  In its spirit, the transformation from the original Schr{\"o}dinger equation to Eq.~(\ref{eq:imp_schr_1}) is similar to the Lee-Low-Pines transformation in momentum space~\cite{lee1953}. In our case, the effective boson-boson interaction
  is hidden in the mixed derivatives in the second term of Eq.~(\ref{eq:imp_schr_1}). 

We look for the real ground state solution $\Phi$ that satisfies the bosonic symmetry, i.e., $\Phi(.., z_i,.. , z_j, ..)=\Phi(.., z_j, .., z_i, ..)$. The bosons are weakly interacting, therefore, we use the product ansatz $\Phi=\prod_{i=1}^N\psi(z_i)$ that gives an approximative solution. We insert this ansatz into Eq.~(\ref{eq:imp_schr_1}), and minimize the energy with respect to $\psi$. This procedure leads to the real Gross-Pitaevski equation (GPE)~\cite{footnote0} for the function $\psi$:

\begin{align}
-\frac{1}{2 \kappa}\frac{\mathrm{d}^2\psi(x)}{\mathrm{d}x^2}+g (N-1) \psi(x)^3  = \mu \psi(x),
\label{eq:GPE}
\end{align}
supplemented by certain boundary conditions at $x=0$ and $x=L$ (see Eq.~(\ref{eq:cond}) below),
$\mu$ is the chemical potential. We write the factor $N-1$ instead of the usual $N$ in front of the $\psi^3$ term as then the equation can be used to obtain an upper bound to the ground state energy also for a small number of particles.  Since we derive the GPE in the rest frame of the impurity, the Bose polaron in our picture is a coherent superposition of the impurity and the condensate that changes dynamically in the vicinity of the impurity. As we show below this viewpoint allows us to perform non-perturbative (in $c$) calculations analytically.

Equation~(\ref{eq:GPE}) can be solved using the Jacobi elliptic functions (see, e.g., Refs.~\cite{carr2000, malomed2000}). The nodeless (in the bulk) real solution we are after reads
\begin{equation}
\psi(\tilde x)=\sqrt{\frac{4K(p)^2 p}{\kappa g L^2\delta^2 (N-1)}}\mathrm{sn}\left(2K(p)\left[\frac{\tilde x}{\delta L}+\frac{1}{2}\right]\bigg|p\right),
\label{eq:wave_func}
\end{equation}
where $\tilde x=x-L/2$, $K(p)$ is the complete elliptic integral of the first kind, and $\mathrm{sn}(x|p)$ is the Jacobi elliptic function~\cite{abram}. The parameters $p \in [0,1)$ and $\delta$ are determined by the boundary conditions and normalization,
\begin{align}
\label{eq:cond}
\int_{0}^{L/2} \psi^2\mathrm{d}x=\frac{1}{2}, \qquad
\frac{\mathrm{d}\psi}{\mathrm{d}x}\bigg|_{x=+0} = c\kappa \psi(0),
\end{align}
where we have used that $\psi(\tilde x)=\psi(-\tilde x)$. 
The corresponding chemical potential and the energy of the impurity are
\begin{align}
\label{eq:eps}
\mu &= 2\frac{p+1}{\kappa \delta^2 L^2}K(p)^2, \\
\epsilon&=\left(\mu-\frac{g (N-1)}{2L}\right)N - g N(N-1)\int_{0}^{L/2} \psi^4(x)\mathrm{d}x. \nonumber
\end{align}
Equations~(\ref{eq:wave_func}), (\ref{eq:cond}) and (\ref{eq:eps}) determine the ground state properties of the system within the mean field approximation for bosons.

\section{Thermodynamic limit} 

We now use these equations
to study the system in the thermodynamic limit, i.e., $N(L)\to\infty$ with $N/L=\rho$, where $\rho$ is the density of the bosons without the impurity. To this end, we note that the parameter $p$ is close to one since $K(p)= \sqrt{\mu \kappa L^2\delta^2}/2\gg 1$, and, thus, the function $\psi$ for $x \in [0,L/2]$ can be written in a much simpler form
\begin{equation}
\psi(\tilde x)\simeq\sqrt{\frac{\mu}{g (N-1)}}\mathrm{tanh}\left(\sqrt{\mu\kappa}L\delta\left[\frac{\tilde x}{\delta L}+\frac{1}{2}\right]\right).
\label{eq:wave_func_thermod}
\end{equation}
The corresponding parameters $\delta$, $\mu$ and $\epsilon$ are
\begin{align}
\label{eq:thermod_d}
\delta &\simeq 1+\frac{2 d}{\sqrt{\gamma\kappa}N}, \qquad d=\frac{1}{2}\mathrm{asinh}\left(\frac{2\rho}{c}\sqrt{\frac{\gamma}{\kappa}}\right),\\
\mu &\simeq \gamma\rho^2\frac{N-1}{N}\left(1-2\frac{\mathrm{tanh}(d)-1}{\sqrt{\gamma \kappa}N}\right), \\
\epsilon&\simeq \frac{\rho^2}{3}\sqrt{\frac{\gamma}{\kappa}}\left[4+\left[-4+\mathrm{sech}^2\left(d\right)\right]\mathrm{tanh}\left(d\right)\right].
\label{eq:energy_thermod}
\end{align}
where $\mathrm{tanh}(x),\mathrm{asinh}(x)$ and $\mathrm{sech}(x)$ are standard hyperbolic functions, and $\gamma\equiv g/\rho$.  Let us discuss the energy $\epsilon$ in more detail. At small values of the impurity-boson coupling $c$ it reads $\epsilon\simeq c\rho$. This result is simply the first order perturbative correction, which follows for any $\gamma$ from the original Hamiltonian $H$ if $V_{ib}$ is treated as a perturbation. Therefore, this expression is applicable as long as $c$ sets the smallest energy scale of the problem. In the opposite limit, i.e., at large values of $c$, we obtain $\epsilon\simeq\rho^2 \sqrt{16\gamma/(9\kappa)}$. This functional dependence follows from the boundary energy of the Lieb-Liniger model~\cite{gaudin1971}, which is reproduced in our case for $\kappa=1$ (infinitely heavy impurity), and the observation that in our equations $\epsilon/\rho^2$ is determined solely by $\sqrt{\gamma/\kappa}$ and $c/\rho$.  Note that this formula overestimates the energy for large values of $\gamma$. In particular, it predicts that $\epsilon\to\infty$ for $\gamma\to\infty$. This prediction is clearly a shortcoming of the mean-field approximation, since we know that for large $\gamma$ the system fermionizes~\cite{girardeau1960} and $\epsilon$ is determined by the chemical potential of a Fermi gas with the same density. 

\begin{figure}
\includegraphics[scale=0.9]{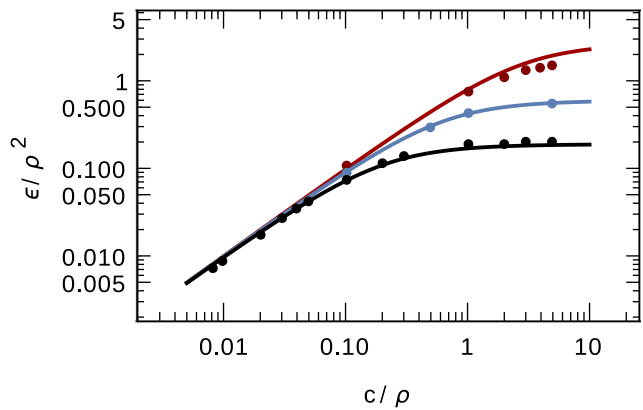}
\caption{The solid lines show the energy of the impurity $\epsilon/\rho^2$ from Eq.~(\ref{eq:energy_thermod}) as a function of $c/\rho$ for $\kappa=1$ (an infinitely heavy impurity), $\gamma=0.02,0.2$ and $4$ (from the bottom to the top). The points are the corresponding results of Ref.~\cite{giorgini2016}.}
\label{fig:termod_analytic1}
\end{figure}

\begin{figure}
\includegraphics[scale=0.9]{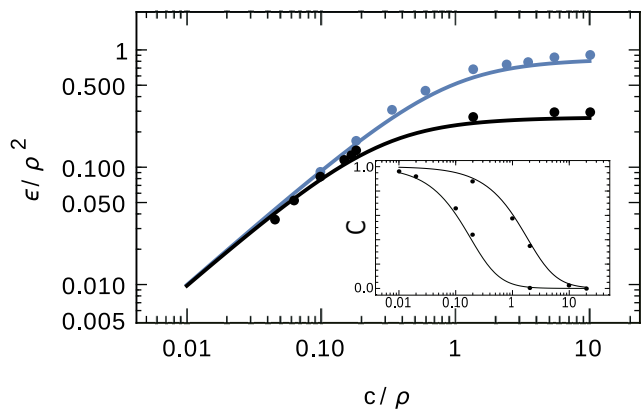}
\caption{The solid lines show the energy of the impurity $\epsilon/\rho^2$ from Eq.~(\ref{eq:energy_thermod}) as a function of $c/\rho$ for $\kappa=1/2$ ($m=M$), $\gamma=0.02$ and $0.2$ (from the bottom to the top). The points are the corresponding results of Ref.~\cite{giorgini2016}. The solid lines in the inset present the contact parameter $C$ defined in Eq.~(\ref{eq:contact}) as a function of $c/\rho$ for $m=M$ and $\gamma=2$ (the upper curve) and $\gamma=0.02$ (the lower curve). The dots are the numerical results of Ref.~\cite{giorgini2016}.}
\label{fig:termod_analytic2}
\end{figure}
 To find the region of applicability of our results we can either estimate effects beyond the GPE or use some numerical results as a reference point. We leave the former approach for a future discussion and focus on the latter. 
To this end, we show in Figs.~\ref{fig:termod_analytic1} and~\ref{fig:termod_analytic2} the quantity $\epsilon/\rho^2$ from Eq.~(\ref{eq:energy_thermod}) together with the recent numerical calculations of Ref.~\cite{giorgini2016}. Note that here only the data points without error bars are included. First, we note that our findings agree well with the results of Ref.~\cite{giorgini2016} for all cases presented. The overall agreement is better for an infinitely heavy impurity, $1/m=0$ (cf.~Fig.~\ref{fig:termod_analytic1}), than for the equal mass case, $m=1$ (cf.~Fig.~\ref{fig:termod_analytic2}). In the former case the results start to deviate noticeably only for $\gamma=4$ at $c/\rho\simeq 4$. As discussed above this deviation is a shortcoming of the mean field approximation, which overestimates $\epsilon$ in this region.   For $m=1$ the results also agree, however, since the relevant interaction parameter within our scheme is $\gamma/\kappa$ the results start to deviate for smaller values of $\gamma$. For this reason we do not plot here the $\gamma=4$ results presented in Ref.~\cite{giorgini2016}.
The comparison to the quantum Monte Carlo calculations suggests that our approach can be used to calculate the energy and structural properties (see below) of these systems for $\gamma/\kappa \lesssim 1$.  For these interactions our analytical expressions for $1/2<\kappa<1$ fill in the gap between the numerical results of Ref.~\cite{giorgini2016}.

Besides the energy, Eqs.~(\ref{eq:wave_func_thermod})-(\ref{eq:energy_thermod}) provide one also the wave function, which in principle allows one to calculate any observable of interest. As an example we have found the density of bosons around the impurity, $
\rho \, \mathrm{tanh}^2(\sqrt{\gamma \kappa}\rho x+d)$, which shows that far from the impurity the bosons are not affected by the impurity and have density $\rho$. We also calculate the contact~$C$~\cite{olschanii2003,tan2008}, which is the density of bosons at the impurity position, $x=0$. 
\begin{equation} 
C\equiv \lim_{L(N)\to\infty} \frac{N \psi^2(x=0)}{\rho}=\mathrm{tanh}^2(d),
\label{eq:contact}
\end{equation}
where $d$ is defined in Eq.~(\ref{eq:thermod_d}).
The parameter $C$ is equal
to the derivative of the energy in Eq.~(\ref{eq:energy_thermod}) with respect to $c$ (recall that $c=-2/a_{1D}$, where $a_{1D}$ is the one-dimensional scattering length).    We plot $C$ in the inset of Fig.~\ref{fig:termod_analytic2} together with the numerical results of Ref.~\cite{giorgini2016}. We see that $C$ decreases from one  to zero as $c$ increases from zero to infinity.  The vanishing of the contact at $1/c=0$ implies that the boson density at the position of the impurity is zero. This is a trivial consequence of the boundary condition~(\ref{eq:bound_cond_initial}) for finite energy solutions.  At $1/c=0$ the density profile of the bosons,
  $\rho \, \mathrm{tanh}^2(\sqrt{\gamma \kappa}\rho x)$, resembles a dark stationary soliton which dresses the impurity. This behavior is already known for a heavy impurity~\cite{carr2000}, but here we show that systems with mobile impurities act similarly. The characteristic length of this soliton is $1/(\rho\sqrt{\gamma \kappa})$. This length becomes larger for smaller values $\gamma$, which imply a higher compressibility  of the gas. Therefore, the polaron in the strongly-interacting regime consists of the impurity and a soliton in the Bose gas. It will be interesting to investigate this correspondence in the future within the presented model for finite interaction strengths in a time-dependent problem.

Another experimentally relevant quantity is the overlap $\mathrm{S}\equiv|\langle \Phi(c=0)|\Phi(c)\rangle|^2=|\langle \psi(c=0)|\psi(c)\rangle|^{2N}$, which determines the probability to populate the interacting ground state after quenching the boson-impurity interaction. We find in our approach:
\begin{equation}
\mathrm{S} = \mathrm{exp}\left(\frac{4\ln(\mathrm{sech} d)-2\mathrm{tanh}d+2+4d-\ln 16}{\sqrt{\gamma \kappa}}\right).
\end{equation}
Using Eq.~(\ref{eq:thermod_d}), 
we see that the overlap is a decaying function of $c$. The largest value is at $c=0$ where $\mathrm{S}=1$. The smallest value $e^{(2-4\ln 2)/\sqrt{\gamma \kappa}}$ is reached at  $1/c=0$. 

\section{Finite $N$} 

We have argued that Eq.~(\ref{eq:GPE}) describes the system well when the number of particles is large. However, this equation can be also used to describe finite number of particles. Note that there is no known (to the best of our knowledge) criterion to determine whether the mean field approximation is applicable for a finite system, therefore, we compare our analytical model with the numerical solution of Eq.~(\ref{eq:imp_schr_1}).  We choose to work with the impenetrable impurity, i.e., $1/c=0$, and $m=M$,  since in the thermodynamic limit it is the most challenging case for the GPE. To have a direct comparison with the previous discussion we fix the density and increase $N$ ($L$). This approach will not only reveal the applicability of the presented method, but also will allow us to study how the energy approaches  its thermodynamic value.

First of all, we note that the $1/c=0$ interaction simplifies the analytical expressions. Indeed, in this case $\delta=1$ and all properties are determined by the value of $p$ alone. It is determined from the equation
\begin{equation}
\frac{4K(p)(K(p)-E(p))}{\kappa \gamma N (N-1)}=1,
\end{equation}
where $E(p)$ is the complete elliptic function of the second kind~\cite{abram}. 
The energy of the impurity is given by
\begin{equation}
\frac{\epsilon}{\rho^2} = \frac{8K^4(p)p+2K^2(p)\kappa \gamma N(N-1)(p+1)}{3\kappa^2\gamma N^2(N-1)}-\frac{\gamma(N-1)}{2}.
\label{eq:energy_finiteN}
\end{equation}
Note that the $\gamma=0$ case leads to $\epsilon=\rho^2\pi^2/(2N\kappa)$, implying that the energy of the impurity goes to zero for $N\to \infty$. This situation is possible due to the high compressibility of the bosons and the absence of an external trap (cf.~Ref.~\cite{amin2015}), which means that the impurity can displace the gas of bosons. 

\begin{figure}
\includegraphics[scale=0.65]{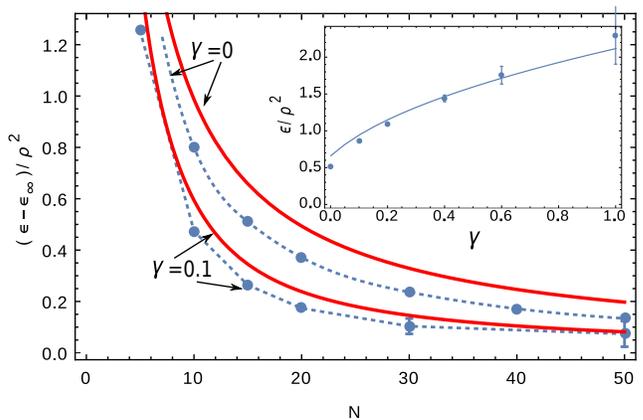}
\caption{The energy of the impurity $\epsilon$ minus the thermodynamic value $\epsilon_{\infty}=\sqrt{16\gamma/(9\kappa)}$ as a function of the particle number $N$, for $\gamma=0.1$ and for $\gamma=0$, in both cases $\kappa=1/2$ ($m=M$) . The solid (red) curves depict Eq.~(\ref{eq:energy_finiteN}). The points are calculated numerically using Eq.~(\ref{eq:floweq}). The dashed lines are to guide the eye. The inset shows $\epsilon$ as a function of $\gamma$ for $N=15$. The solid line is from Eq.~(\ref{eq:energy_finiteN}), points are the numerical results. %, and dotted line shows $\epsilon_{\infty}$.
}
\label{fig:imp_energy}
\end{figure}

 To investigate the Schr{\"o}dinger equation numerically, we use the flow equation method for bosonic systems presented in detail in Ref.~\cite{volosniev2017}. Here, we use it to solve Eq.~(\ref{eq:imp_schr_1}), which does not contain the coordinate of the impurity anymore. In this secton, we consider an impenetrable impurity. However, we believe that the flow equations of Ref.~\cite{volosniev2017} can give accurate results also for finite values of $c$.  In this method the parameters of the Hamiltonian in second quantization are assumed to 'evolve' with the flow parameter $s$, such that
\begin{equation}
H(s)=\sum_{i,j} h^{(1)}_{ij}(s) a_i^\dagger a_j + \frac{1}{2}\sum_{i,j,k,l}h^{(2)}_{ijkl}(s) a_i^\dagger a_j^\dagger a_k a_l, 
\end{equation}
 where $a_i$ ($a_i^\dagger$) is the bosonic creation (annihilation) operator and the initial condition is $H(0)=H$. During the flow the couplings to the ground state decrease and when the parameter $s$, which can be thought of as a resolution scale, is large the ground state is decoupled and its energy is easily obtained. 

The flow is described by the system of differential 
equations~(see, e.g., Ref.~\cite{kehrein2006})
\begin{equation}
\frac{\mathrm{d}H(s)}{\mathrm{d}s}=[\eta(s),H(s)],
\label{eq:floweq}
\end{equation}
with $\eta$ the antihermitian operator written as
\begin{align}
\eta(s)&=\sum_{i,j}\eta^{(1)}_{ij}(s)a_i^\dagger a_j + \frac{1}{2} \sum_{i,j,k,l}\eta^{(2)}_{ijkl}(s)a_i^\dagger a_j^\dagger a_k a_l.
\end{align} 
The parameters  $\eta_{ij}^{(1)}$ and $\eta_{ijkl}^{(2)}$ should be chosen such that the flow eliminates the couplings of some reference state (that ideally contains our preliminary knowledge of the ground state) to the other states, see below.
The commutator in Eq.~(\ref{eq:floweq}) contains also three-body operators, they must be truncated in order for our scheme to work. To this end, we use the basis\cite{footnote} $\{\mathrm{sin}(\pi i z_i/L)\}$ to construct matrix representations of operators, and neglect the operators that excite three particles simultaneously from our reference state $\prod \mathrm{sin}(\pi z_i/L)$. The operator $\eta$ can be chosen in various ways (see, e.g., \cite{hergert2016}). We construct $\eta$ from the piece of $H$ that should be eliminated, i.e., $\eta_{ijkl}=h^{(2)}_{ijkl}\delta_{k0}\delta_{l0}$ etc, where $\delta_{ij}$ is the Kronecker delta. For our problem, the operator $\eta$ generates the flow that at $1/s=0$ decouples the reference state from the rest, giving us an approximation to the ground state energy.  Because we truncate the flow equations at the level of three-body operators and beyond, the results are not exact. The accuracy can be estimated using the neglected pieces, see Ref.~\cite{volosniev2017}. We assess them and plot as the error bars in Fig.~\ref{fig:imp_energy}. The accuracy worsens when the number of particles or the boson-boson interaction increases. However, for the most considered cases the results are essentially exact, hence they can be used to check the validity of the analytical model.

We present our findings in Fig.~\ref{fig:imp_energy} for $\gamma=0$ and $\gamma=0.1$. We see that the energies in both cases slowly converge to their thermodynamic values, denoted as $\epsilon_{\infty}\equiv\rho^2\sqrt{16 \gamma/(9\kappa)}$, from Eq.~(\ref{eq:energy_thermod}). 
 The numerical results shown as dots agree
reasonably well with the analytical formula for all considered cases, but there are some deviations for small particle numbers, where the neglected few-body correlations are important.  These deviations are more pronounced at weak interaction strengths, see the inset where we show the dependence of $\epsilon$ on $\gamma$ for $N=15$. 
Finally, we note that the rate of convergence to the thermodynamic limit is relatively slow for small values of $\gamma$. We attribute this behavior to a high compressibility of the bosonic gas, which, in particular, implies that to realize the thermodynamic limit in a lab one needs a very low concentration of the impurity atoms.  Note that if the boson-boson interactions were strong the dynamics would be different -- even a few majority atoms would be able to form a many-body enviroment for the impurity (cf. Ref.~\cite{selim2013} for fermions).

\section{Summary} 

We have presented a simple analytical model of an impurity in a one-dimensional Bose gas. 
Within this model, we have derived the ground state energy and showed that it agrees with numerical results for moderate and weak boson-boson interaction strengths. The model also allowed us to get insight into structural properties of the system, such as the contact parameter.  For the mass-balanced case, together with the exact solution available for strongly-interacting
bosons it gives a complete analytical picture of the Bose polaron problem in one spatial
dimension, both in the thermodynamic limit and for
systems with a finite particle number.
We hope that our new method will provide further novel insights into the Bose polaron problem in 1D.  In particular,
it would be interesting to utilize our method to study attractive boson-impurity interactions (i.e., $c<0$), to explore the Bose polaron problem in higher spatial dimensions, and to investigate the evolution of the system after a quench of the boson-impurity interaction on experimentally relevant time scales. Moreover, our method produces an accurate reference state, which can be used as a starting point in various numerical approaches, e.g., in the flow equation method used here~\cite{volosniev2017}.

\vspace*{2em}

\begin{acknowledgments}
        { We thank Nikolaj Zinner for many inspiring discussions and comments on the manuscript.  We also thank Gregory Astrakharchik and the participants of EMMI Workshop {\it "From few to many: Exploring quantum systems one atom at a time"} for useful conversations.  A.~G.~V. gratefully acknowledges the support of the Humboldt Foundation. H.-W. H. was supported in part by the Deutsche Forschungsgemeinschaft through SFB 1245 and
          by the German Federal Ministry of Education and Research under contract 05P15RDFN1.}
\end{acknowledgments}

\section*{Appendix}
In this Appendix, we demonstrate that the solution $\Phi(z_1,...z_N)$ of Eq.~(\ref{eq:imp_schr_1})
solves the original Schr{\"o}dinger equation, i.e, we show that 
\begin{equation}
\left(-\frac{1}{2m}\frac{\partial^2}{\partial y^2}-\frac{1}{2}\sum_{i=1}^N\frac{\partial^2}{\partial x_i^2}\right)\Phi(z_1,...,z_N) = \varepsilon_{gr}\Phi(z_1,...,z_N),
\label{eq:app1}
\end{equation}
for every ordering of the particles, and that $\Phi$ satisfies the boundary conditions associated with the interactions and the geometry. 
To this end, we note that according to the prescription $z_i=L\theta(y-x_i)+(x_i-y)$ the derivatives for every ordering read
\begin{equation}
\frac{\partial}{\partial y}=-\sum_i \frac{\partial}{\partial z_i}, \qquad
\frac{\partial}{\partial x_i}=\frac{\partial}{\partial z_i}.
\label{eq:app2}
\end{equation}
Using these equations, we immediately obtain Eq.~(\ref{eq:imp_schr_1}) from Eq.~(\ref{eq:app1}).
As a consequence, $\Phi$ is a solution of Eq.~(\ref{eq:app1}) by construction.  Let us now consider the boundary conditions associated with the boson-impurity interaction $\delta(x_i-y)$. For an eigenstate $\Psi$ of the Hamiltonian~(\ref{eq:ham1}), we write the
boundary conditions as
\begin{eqnarray}
\left(m\frac{\partial}{\partial x_i}-\frac{\partial}{\partial y}\right)^{x_i=y^+}_{x_i=y^-}
\Psi&=&2cm\Psi(x_i=y), \label{eq:app3}\\
 \Psi(x_i=y^+)&=&\Psi(x_i=y^-).
\label{eq:app4}
\end{eqnarray}
Using Eq.~(\ref{eq:app2}) in Eqs.~(\ref{eq:app3}) and~(\ref{eq:app4}),  we obtain the conditions on $\Phi$ from
Eq.~(\ref{eq:bound_cond_initial}), which are therefore satisfied by construction. The validity of other boundary conditions can be proven in a similar manner. Finally, we note that $\Phi$ is also an eigenstate of the total angular momentum operator $\frac{\partial}{\partial y}+\sum_{i}\frac{\partial}{\partial x_i}$ with zero eigenvalue. This implies that our transformation singles out the
manifold of zero total angular momentum where we expect the ground state to be.


\begin{thebibliography}{99}
\bibitem{footnote0}  Note that the GPE in the new coordinates has this simple form only for real solutions. If complex solutions and flux phenomena are considered it is more complicated.
\bibitem{footnote} This is a complete basis in which the impurity is impenetrable.
\end{thebibliography}
 \end{document}